\documentclass[a4paper,11pt]{article}
\usepackage{mathrsfs,amsmath, amssymb}
\usepackage{a4wide}
\usepackage{bm}
\usepackage{graphicx}
\usepackage{amsfonts}
\usepackage{upgreek}

\newcommand{\bmN}{\bm{\nabla}}
\newcommand{\pfrac}[2]{\frac{\partial{#1}}{\partial{#2}}}
\newcommand{\refe}[1]{(\ref{#1})}
\newcommand{\bmDot}{\bm{\cdot}}
\newcommand{\bmtau}{\bm{\tau}}
\newcommand{\tbU}{\vect{u}}
\newcommand{\tbN}{\vect{n}}
\newcommand{\ilfrac}[2]{{#1}/{#2}}
\newcommand{\pilfrac}[2]{\partial_{#2} {#1}}
\newcommand{\ppilfrac}[2]{\partial^2_{#2} {#1}}
\newcommand{\lr}[1]{\left( {#1} \right)}
\newcommand{\lrsq}[1]{\left[ {#1} \right]}
\newcommand{\lrcur}[1]{\left\{ {#1} \right\}}
\newcommand{\lreval}[1]{\left. {#1} \right|}
\newcommand{\infinity}{\infty}
\newcommand{\Ca}{\operatorname{Ca}}
\newcommand{\Cn}{\operatorname{Cn}}
\newcommand{\td}[1]{\tilde{#1}}
\newcommand{\eg}{\textit{e.g.}}
\newcommand{\ie}{\textit{i.e.}}
\newcommand{\inin}[1]{\td{#1}}

\newcommand{\upd}{\textrm{d}}
\newcommand{\tens}[1]{\textbf{#1}}
\newcommand{\vect}[1]{\textbf{#1}}

\begin{document}
\title{On the moving contact line singularity: Asymptotics of a diffuse-interface model}
\author{David N. Sibley$^1$, Andreas Nold$^1$, Nikos Savva$^{2,1}$, and Serafim Kalliadasis$^1$
\vspace{0.5cm}
\\
\small{$^1$ Department of Chemical Engineering, Imperial College London, London SW7 2AZ, UK}\\
\small{$^2$ School of Mathematics, Cardiff University, Senghennydd Road, Cardiff CF24 4AG, UK}
}
\date{Published: Eur. Phys. J. E (2013) \textbf{36}: 26, DOI: 10.1140/epje/i2013-13026-y}

\maketitle
\begin{abstract}
The behaviour of a solid-liquid-gas system near the three-phase contact line
is considered using a diffuse-interface model with no-slip at the solid and
where the fluid phase is specified by a continuous density field. Relaxation
of the classical approach of a sharp liquid-gas interface and careful
examination of the asymptotic behaviour as the contact line is approached is
shown to resolve the stress and pressure singularities associated with the moving contact line problem.
Various features of the model are scrutinised, alongside extensions to incorporate slip, finite-time
relaxation of the chemical potential, or a precursor film at the wall.
\end{abstract}

%
%
%

\section{Introduction}

A moving contact line occurs
at the location
where two ostensibly
immiscible fluids and a solid meet. It arises in a wide range of both natural
and technological processes, from insects walking on water~\cite{insect} and
the wetting properties of plant leaves~\cite{leaves} to
coating~\cite{coating}, inkjet printing~\cite{inkjet,inkjet2} and oil
recovery~\cite{oil}. In addition to its crucial role in wide-ranging
applications, it remains a persistent problem, a long-standing and
fundamental challenge in the field of fluid dynamics, despite its apparent
simplicity at first sight (see \eg~review
articles~\cite{Dussan79,deGennesrev,blake2006physics,BonnEggers}). Not
surprisingly it has been investigated extensively, both experimentally and
theoretically, for several decades.

At the heart of the moving contact line problem is that, when treated classically as
two immiscible fluids moving along a solid surface satisfying the no-slip
condition, there is no solution due to the multivalued velocity at the
contact line,~\cite{DussanDavis,ShikhSingualarities06}. This is known most famously
in the literature
as
the problem of a non-integrable stress singularity, a result published a few
decades ago along with the nonphysical prediction that an infinite force is
required to submerge a solid object~\cite{HuhScriv71}.

The resolution of the problem may have initially appeared straightforward.
The no-slip condition at the wall could not be satisfied, thus some form of
slip in the contact line vicinity should be allowed. Navier-slip, written down in the
early 19th Century~\cite{Navier}, was a prime candidate, a form of which was
suggested in the concluding remarks of~\cite{HuhScriv71}. The fact that
wetting and the moving contact line remain an open debate and a fruitful research area is
largely due to the particular microscale ingredients that may
alleviate the problem being numerous and hotly debated---see \eg~the wide
range of discussion articles recently,~\cite{epj_entire}. Various alternative
models to slip at the wall include: a precursor film ahead of the
contact line~\cite{SchwartzEley}; rheological effects~\cite{WeidnerSchwartz}; treating
surfaces as separate thermodynamic entities with dynamic surface
tensions~\cite{Shikh93} (see a recent critical investigation of this
model,~\cite{My_Shikh}); introducing numerical slip~\cite{RenardySlip};
including evaporative fluxes~\cite{ColinetRednikov};
and considering the interface to be diffuse, numerical work reviewed \eg~in~\cite{anderson_rev}.

Here, we examine analytically a diffuse-interface model
without any other ingredients, being both self-consistent and
physically relevant: rather than considering a sharp fluid-fluid
interface as a surface of zero thickness where quantities
(\eg~density) are, in general, discontinuous, it considers the
interface to have a non-zero, finite, thickness with quantities varying
smoothly but rapidly, in agreement with developments from the
statistical mechanics of liquids community (\eg~\cite{EvansReview,yatsyshin1}).
The fluid density $\bar\rho$ then acts
as an order parameter such that in the sharp-interface limit the two
bulk fluids satisfy $\bar\rho=\rho_L$ and $\bar\rho=\rho_V=0$, being
liquid and vapour respectively, where we consider the behaviour of
the system with vapour phase of negligible density,
noting that an equivalent double-well free energy form (to be described later,
in sect. \ref{sec:ps}), with zero bulk vapour density, is also used by
Pismen and Pomeau~\cite{PismenPomeau} and is physically relevant for liquid-gas systems.
The interface between liquid and gas may then be defined as the location where
$\bar\rho=(\rho_L+\rho_V)/2$.

Diffuse-interface models have been popular for numerical
implementation as tracking of the fluid-fluid interface is not
required in the resulting free-boundary problem, instead the
interface is inferred from density field contours. For
solid-liquid-gas systems the seminal study of
Seppecher~\cite{seppecher} is often referred to when suggesting that
diffuse-interface models resolve the moving contact line problem.
Whilst Seppecher's work contains some discussion of the asymptotics, the
analysis was largely incomplete, with asymptotic regions being
probed without careful justification and the crucial behaviour close
to the contact line only investigated numerically (a number of
constraints were also imposed, \eg~90$^\circ$ contact angles and
fluids of equal viscosity). Full numerical simulations for the
liquid-gas problem have also been undertaken
(\eg~in~\cite{BriantYeomansEarly,BriantYeomans1}); binary fluids have also been examined using diffuse-interface methods of a different
form, where a coupled Cahn--Hilliard equation models the diffusion
between the two
components~\cite{KhatavkarJFM,DingSpeltJFM,YueZhouFeng}.

Here, we undertake an analytical investigation by considering the
contact line behaviour for a liquid-gas system with two basic
elements: (a) the interface has a finite thickness, which is
expected from statistical mechanics studies as noted earlier, and
(b) the no-slip condition is applied at the wall. This
diffuse-interface model then resolves the moving contact line
problem without the need to model any further physical effects
from the microscale. An important
ramification of this analysis is that the wetting boundary condition
used in conjunction with diffuse-interface in existing numerical
studies of liquid-gas systems needs to be appropriately modified,
otherwise it leads to a density gradient orthogonal to the wall at large distances from the contact
line. The possibility of density variations such as these are often
included when disjoining pressure models are considered, where many
studies utilise the long-wave (or lubrication) approximation,
\eg~\cite{deGennesrev,oron1997long}. 
In the 1985 review of de Gennes \cite{deGennesrev}, the thickness of precursor films was discussed. 
For complete wetting on a dry substrate, nanometric films, decaying ahead
of the contact line, were predicted. A recent study of intermolecular forces in the contact line region using approaches from statistical mechanics, namely density-functional theory~\cite{antoniojfm}, demonstrated that for the case of partially wetting fluids, a constant-thickness nanometric film of a few molecular diameters is adsorbed in front of the contact line. More recent experiments have also been performed (\eg~to probe the dynamics of such nanoscale films \cite{KavPre11}).

In the macro/mesoscopic setting of this work, and to isolate the contribution of the diffuse-interface model
without depletion/enrichment near the solid substrate, we thus predominately consider the case where the bulk
densities are valid up to the walls (but other cases are
considered), and we make no assumptions on thin films---rather
considering arbitrary finite contact angles.

\begin{figure}[t]
\centering
	\includegraphics{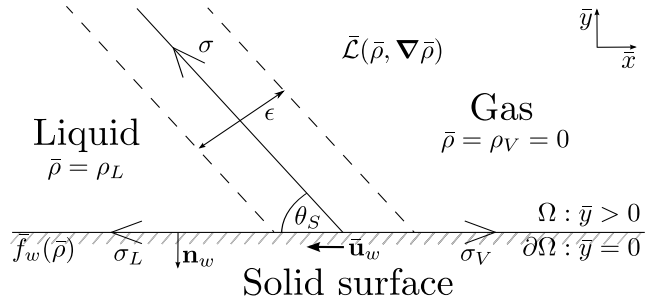}\
	\caption{Sketch of the diffuse-interface model near a wall.}
\label{fig:diffinterface}
\end{figure}

\section{Problem specification}
\label{sec:ps}
Consider a fluid in the upper half $(\bar x,\bar y)$-plane~$\Omega$ with a
solid surface~$\partial\Omega$ at $\bar y=0$; see
Fig.~\ref{fig:diffinterface}. The free energy of the system has contributions
from an isothermal fluid with a Helmholtz free energy functional and from a
wall energy $f_w=f_w(\bar\rho)$, thus given by $\mathscr{F} = \int_\Omega
\bar{\mathcal{L}} \, \upd\Omega + \int_{\partial\Omega} f_w \, \upd A$, where
\begin{equation}
 \bar{\mathcal{L}} = \bar\rho \bar{f}(\bar\rho) + {K}|\bmN\bar\rho|^2/{2} -\bar{G}\bar\rho, \label{eq:origL}
\end{equation}
with $\bar{G}$ the chemical potential, $K$ a gradient energy
coefficient (assumed constant for simplicity), and $\bar\rho \bar
f(\bar\rho)$ a double well potential chosen to give the two
equilibrium states $\bar\rho=\{0,\rho_L\}$. Such a form for the free
energy and associated diffuse-interface approximations has been used
by numerous authors for wetting problems such as Seppecher
\cite{seppecher}, Pismen and Pomeau \cite{PismenPomeau}, and Pismen
\cite{Pismenmeso}; see also the review by de~Gennes
\cite{deGennesrev}. The effect of the non-local terms neglected in
the local approximation to obtain such a free energy has been
considered at equilibrium in the aforementioned studies
\cite{Pismenmeso,antoniojfm}, where the long-range intermolecular
interactions are responsible for an algebraic decay of the density
profile away from the interface instead of the exponential one as
predicted here (seen later, in \refe{eq:cak0}). In our dynamic
situation, we focus on the local approximation to elucidate the
contact line behaviour in a simplified, yet widely used setting. The
density field augments the usual hydrodynamic equations via the
capillary (or Korteweg) stress tensor $\bar{\tens{T}}$ through
\begin{equation}
 \bar f(\bar\rho) = \frac{K}{2\epsilon^2}\bar\rho\lrsq{ 1-\frac{\bar\rho}{\rho_L} }^2, \quad \bar{\tens{T}} = \bar{\mathcal{L}}\tens{I} - \bmN \bar\rho \otimes \pfrac{\bar{\mathcal{L}}}{(\bmN \bar\rho)},\label{eq:origf}
\end{equation}
where $\tens{I}$ is the identity tensor, and with $\epsilon$ being the interface width, and $\bar{\tens{T}}$ arising from
Noether's theorem,~\cite{anderson_rev}. 
Following \cite{seppecher}, \cite{anderson_rev} and \cite{PismenPomeau},
using \refe{eq:origL} and coupling in
the compressible Stokes equations (assuming creeping flow) the capillary
tensor with the usual viscous stress tensor $\bar\bmtau$, taken as deviatoric
for simplicity, yields
\begin{gather}
 \pilfrac{\bar\rho}{\bar{t}} + \bmN\bmDot(\bar\rho\bar\tbU) = 0, \qquad \bmN \bmDot (\bar{\tens{T}} + \bar\bmtau) =0, \nonumber \\
 \bar{\tens{T}} = \left( \bar\rho \bar f(\bar\rho) + {K}|\bmN\bar\rho|^2/2 - \bar{G}\bar\rho \right)\tens{I} - K\bmN\bar\rho\otimes\bmN\bar\rho, \nonumber \\
 \bar\bmtau = \bar\mu(\bar\rho)\left[(\bmN\bar\tbU)+(\bmN\bar\tbU)^\mathrm{T} - {2}(\bmN\bmDot\bar\tbU)\tens{I}/3\right], \nonumber\\
 \bar{G} = -K\nabla^2\bar\rho + \pilfrac{}{\bar\rho}(\bar\rho \bar f(\bar\rho)),\label{eq:gesdim}
\end{gather}
where $\bar\tbU$ and $\bar\mu(\bar\rho)$ are the fluid velocity and
viscosity, respectively, $\bar{t}$ is time, and the thermodynamic pressure is
given by $\bar p=\bar\rho^2 \bar f'(\bar\rho)$. The form for $\bar{G}$ arises
from the Euler-Lagrange equation corresponding to the free energy, and we
take $\bar\mu(\bar\rho) = \mu_L \bar\rho/\rho_L$, giving the viscosities for
the two equilibrium states as $\mu_L$ and $\mu_V = 0$.

\begin{figure}[t]
\centering
	\includegraphics{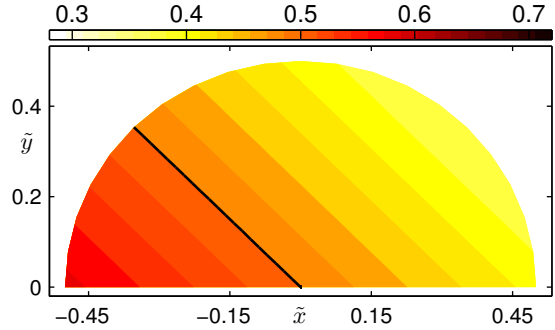}\
	\caption{The equilibrium density behaviour (contours) near the contact line for $\theta_S=\pi/4$, from eq.~\refe{eq:cak0} in inner variables.}
\label{fig:eqd}
\end{figure}

On the solid surface $\partial\Omega$, with normal $\tbN_w$, we impose
\begin{equation}
 \bar{\tbU} = \bar{\tbU}_w, \qquad
 K \tbN_w \bmDot \bmN \bar\rho + \bar f'_w(\bar\rho) = 0,\label{eq:origbcs}
\end{equation}
with wall velocity $\bar{\tbU}_w=(-V,0)$ in Cartesian coordinates. The first
condition is classical no-slip, whilst the second arises by variational
arguments and is termed the \textit{natural} (or \textit{wetting}) boundary
condition, \cite{YueZhouFeng04}. $\bar f_w(\bar\rho)$ is chosen to satisfy Young's law at
the contact line, with solid-liquid, solid-gas and liquid-gas surface tensions $\bar
f_w(\rho_L)=\sigma_L$, $\bar f_w(\rho_V)=\sigma_V$ and $\sigma$ respectively,
and with contact angle $\theta_S$. A cubic is the lowest-order
polynomial required for the wall free energy to be minimised by the
bulk densities, to prevent depletion/enrichment away from the contact line, \ie~$\bar
f'_w(\bar\rho_L)=\bar f'_w(0)=0$. Whilst cubic forms are used for binary fluid
problems, \eg~\cite{jacqmin,YueZhouFeng}, this is unlike the linear forms proposed in
previous studies for liquid-gas problems
\cite{seppecher,BriantYeomansEarly,BriantYeomans1,qianliqgas}, and allows us to
consider a diffuse-interface model without further physical effects from the microscale (although this is relaxed in the following section). We define
\begin{equation*}
\bar f_w(\bar\rho) = \lrsq{{\rho_L^{-3}\sigma\cos\theta_S} ( 4\bar\rho^3 -6\bar\rho^2\rho_L + \rho_L^3 ) + \sigma_V+\sigma_L }/2,
\end{equation*}
giving 
\begin{equation*}
\bar f'_w(\bar\rho) = -{6\sigma}\bar\rho\left( \rho_L - \bar\rho
\right)\cos\theta_S/{\rho_L^3}, 
\end{equation*}
and $\bar f_w(0)-\bar f_w(\rho_L) =
\sigma\cos\theta_S$, with Young's law thus satisfied. 
It is noteworthy that
\refe{eq:origbcs}(b) may be replaced by a constant density
condition if a precursor film/disjoining pressure model is to be considered,
\ie~$\bar{\rho}=\rho_a$ on $\partial\Omega$ (as used in
\cite{PismenPomeau}, and considered in the following section).
Finally, for a one-dimensional density profile $\bar\rho(z)$ in
equilibrium (\ie~with $\bar{G}=0$ required by our choice of $\bar\rho\bar{f}(\bar\rho)$ to have equally stable bulk fluids, and $\bar{\tbU}=0$), the surface tension across the
interface is
\begin{equation}
 \sigma = K \int_{-\infinity}^\infinity\left( \frac{\upd\bar{\rho}}{\upd\bar{z}} \right)^2\, \upd\bar{z}=\frac{K\bar\rho_L^2}{6\epsilon}\label{eq:dimsurfaceT},
\end{equation}
see \eg~\cite{CH58}, and we note Eqs. \refe{eq:gesdim}--\refe{eq:origbcs}, without the specific choices for $\bar{\mu}(\bar\rho)$, $\bar{f}(\bar\rho)$  and $\bar{f}_w(\bar\rho)$ are as derived/used in \cite{seppecher,anderson_rev,PismenPomeau}.
To nondimensionalise, we use typical length, velocity and density scales $X$,
$V$ and $\rho_L$ respectively. The pressure and viscous stress are scaled
with $\ilfrac{\mu_L V}{X}$, and the capillary stress with
$\ilfrac{K\rho_L^2}{(\epsilon X)}$. Finally, $\bar{G}$ is scaled with
$\ilfrac{K\rho_L}{(\epsilon X)}$ and $\bar{f}$ with
$\ilfrac{K\rho_L}{\epsilon^2}$. The governing equations then contain the nondimensional parameters
\begin{equation*}
 \Cn = \ilfrac{\epsilon}{X}, \quad\mbox{and}\quad \Ca_k = \ilfrac{\mu_L V \epsilon}{(K \rho_L^2)},
\end{equation*}
being the Cahn number and a modified Capillary number based on the model
parameter $K$, respectively. Nondimensional variables are denoted as their
dimensional counterparts with bars dropped, \eg~nondimensional bulk
densities are $\rho=\{0,1\}$. $\Ca_k$ is related to the usual Capillary
number, $\Ca=\mu_L V / \sigma = 6\Ca_k$, through \refe{eq:dimsurfaceT}.
The governing equations \refe{eq:gesdim} in nondimensional form are
\begin{gather}
 \pilfrac{\rho}{t} + \bmN\bmDot(\rho\tbU) = 0, \quad
 \tens{M} =  {\Ca_k}^{-1}\tens{T} + \bmtau, \quad \bmN \bmDot \tens{M} = 0, \nonumber\\
 \tens{T} = \left( {\Cn}^{-1} \rho f(\rho) + {\Cn}|\bmN\rho|^2/2 - G\rho \right)\tens{I} -\Cn\bmN\rho\otimes\bmN\rho, \nonumber\\
 \bmtau = \rho\lrsq{(\bmN\tbU)+(\bmN\tbU)^\mathrm{T} - {2}(\bmN\bmDot\tbU)\tens{I}/{3}}, \nonumber\\
 G = -\Cn\nabla^2\rho + {\Cn}^{-1}\pilfrac{}{\rho}(\rho f(\rho)),\label{eq:nd1}
\end{gather}
where $\tens{M}$ is introduced as the total stress tensor, $p=(\Cn\Ca_k)^{-1}\rho^2f'(\rho)$, and from \refe{eq:origf}:
\begin{align}
 f(\rho) = \frac{\rho}{2}(1-\rho)^2, \quad \mbox{giving} \quad p = \frac{\rho^2(1-3\rho)(1-\rho)}{2\Cn\Ca_k}.\label{eq:nd2}
\end{align}
On the solid surface $\partial\Omega$, from \refe{eq:origbcs}, we have
\begin{align}
 \tbU = \tbU_w, \qquad
 \Cn \tbN_w \bmDot \bmN \rho = \cos\theta_S(1-\rho)\rho ,\label{eq:ndge2}
\end{align}
where $\tbU_w=(-1,0)$, and $\tbN_w=(0,-1)$, in Cartesian components.  We can rewrite the governing equations \refe{eq:nd1} for steady flows as
\begin{align}
\bmN\bmDot(\rho\tbU) &= 0, \nonumber\\
\bmN p &= (2\Cn\Ca_k)^{-1}\bmN [\rho^2(1-3\rho)(1-\rho)],\nonumber\\
0&=\Ca_k^{-1} {\Cn} \ \rho\bmN(\nabla^2\rho) - \bmN p  \nonumber\\ &\qquad + \rho[\bmN^2\tbU + \bmN(\bmN\bmDot\tbU)/3]
\nonumber\\
& \qquad + [(\bmN\tbU)+(\bmN\tbU)^\mathrm{T} - 2(\bmN\bmDot\tbU)\tens{I}/3] \bmN\rho
,\label{eq:ndge1}
\end{align}
and the boundary conditions on $\partial\Omega$ remain as above, in \refe{eq:ndge2}.
We initially consider the equilibrium behaviour of the system, corresponding to $\Ca_k\ll1$, 
to provide a basis for comparison when the dynamic behaviour is analysed,
where eqs.~\refe{eq:ndge2}--\refe{eq:ndge1} are thus reduced to
\begin{equation*}
2{\Cn^2}\rho\bmN(\nabla^2\rho) = \bmN \lrsq{\rho^{2}(1-3\rho)(1-\rho)},
\end{equation*}
subject to $-\Cn\pilfrac{\rho}{y} = \cos\theta_S(1-\rho)\rho$ at $y=0$, and with $\rho\to\{0,1\}$ and $\nabla^2\rho\to0$ as $x\to\pm\infinity$ to obtain the expected bulk behaviour. The solution subject to the above conditions is
\begin{equation}
 \rho = \lr{1-\tanh\lrsq{(2\Cn)^{-1}(x\sin\theta_S+y\cos\theta_S) }}/2, \label{eq:cak0}
\end{equation}
having also fixed the interface at $\rho=1/2$. This profile is planar and
at angle $\theta_S$, shown in Fig.~\ref{fig:eqd} in inner variables
(where $\{x,y\} = \Cn \{\td{x},\td{y}\}$, for comparison to forthcoming plots).

For physical systems, the scale over which the density varies between liquid and gas
is much smaller than the macroscopic length scale, and hence $\Cn\ll1$. The asymptotic behaviour
as $\Cn\to0$ is known as the sharp-interface limit, and understanding of it is of vital importance
when considering diffuse-interface models as classical continuum models should be recovered if
correct predictions are to be found.

A careful asymptotic analysis of the outer solution away from the interface,
and of the interfacial region away from the wall (using body fitted
coordinates), shows that the expected sharp-interface equations (the Stokes
equations with no-slip and the usual capillary surface stress conditions) are indeed
recovered. 

Consider now the inner region near to both interface and wall in polar
coordinates with $r=O(\Cn)$. The scaling $\bmN = \Cn^{-1}\inin{\bmN}$ (\ie~$r=\Cn\inin{r}$, with inner variables
denoted with tildes) retains all terms in the governing equations and
boundary conditions \refe{eq:ndge2}--\refe{eq:ndge1}, giving a complete dominant balance. The steady governing equations, from \refe{eq:ndge1}, in this inner region are thus
\begin{align}
\inin\bmN\bmDot(\inin\rho\inin\tbU) = 0,\label{eq:ndinge1}
\end{align}
and
\begin{align}
&\Ca_k^{-1} \inin{\rho}\inin\bmN(\inin\nabla^2\inin\rho) - (2\Ca_k)^{-1}\inin\bmN [\inin\rho^2(1-3\inin\rho)(1-\inin\rho)] \nonumber\\ &
\qquad\qquad + \inin\rho[\inin\bmN^2\inin\tbU + \inin\bmN(\inin\bmN\bmDot\inin\tbU)/3]
\nonumber\\
&  \qquad\qquad + [(\inin\bmN\inin\tbU)+(\inin\bmN\inin\tbU)^\mathrm{T} - 2(\inin\bmN\bmDot\inin\tbU)\tens{I}/3] \inin\bmN\inin\rho
=  0,\label{eq:ndinge2}
\end{align}
and on the solid surface $\partial\Omega$, the boundary conditions are
\begin{align}
 \inin\tbU = \inin\tbU_w, \qquad
 \tbN_w \bmDot \inin\bmN \inin\rho = \cos\theta_S(1-\inin\rho)\inin\rho ,
\end{align}
where in polar coordinates, and with $\inin{u}$ and $\inin{v}$ being the radial and angular velocity components, these reduce to
\begin{align}
 \inin{u} &= -1, & \inin{v} &= 0, & -\frac{1}{\inin{r}}\pfrac{\inin{\rho}}{\theta} &= \cos\theta_S(1-\inin{\rho})\inin{\rho},\label{eq:ndinge3} 
\intertext{on $\theta=0$, and}
\inin{u} &= 1, & \inin{v} &= 0, & \frac{1}{\inin{r}}\pfrac{\inin{\rho}}{\theta} &= \cos\theta_S(1-\inin{\rho})\inin{\rho},\label{eq:ndinge4}
\end{align}
on $\theta=\pi$. Of particular
interest is the behaviour as the contact line is approached---the location where a
stress singularity or no solution (due to multivalued velocity) arises in the
classical formulation of the problem.
To consider the asymptotic solution as the contact line is approached, we expand
\begin{equation*}
 \{\inin\rho,\inin{u},\inin{v}\} = \sum_{i=0}^\infty \{ \inin\rho_i(\theta), \inin{u}_i(\theta), \inin{v}_i(\theta) \} \, \inin{r}^i,
\end{equation*}
in \refe{eq:ndinge1}--\refe{eq:ndinge2},
and find at leading order
\begin{equation*}
 {\inin{\rho}_0}(\inin{u}_0 + \inin{v}_0') =  - {\inin{\rho}_0'}\inin{v}_0, \qquad
 \inin{\rho}_0''' = 0, \qquad
 \inin{\rho}_0'' = 0,
\end{equation*}
subject to $\inin{\rho}_0'=0$ on $\theta=\{0,\pi\}$, from \refe{eq:ndinge3}--\refe{eq:ndinge4}. The density is solved as
$\inin{\rho}_0=1/2$, having imposed its expected value at the interface. To
find the leading-order velocities, we continue to first order in the
governing equations \refe{eq:ndinge1}--\refe{eq:ndinge2}, where
\begin{gather*}
 \inin{u}_0 = -\inin{v}_0', \qquad
 \Ca_k(\inin{v}_0''' + \inin{v}_0') + \inin{\rho}_1'' + \inin{\rho}_1 = 0, \nonumber\\
 \Ca_k(\inin{v}_0'' + \inin{v}_0) - \inin{\rho}_1''' - \inin{\rho}_1' = 0.
\end{gather*}
with the wetting and no-slip conditions from \refe{eq:ndinge3}--\refe{eq:ndinge4} being
\begin{align*}
 \inin{u}_0 &= -1, & \inin{v}_0 &= 0, & -\inin{\rho}_1' &= \cos(\theta_S)/4,
\intertext{on $\theta=0$, and}
\inin{u}_0 &= 1, & \inin{v}_0 &= 0, & \inin{\rho}_1' &= \cos(\theta_S)/4,
\end{align*}
on $\theta=\pi$. We also assert that the profile must
be planar at these very small distances to the contact line for a well-defined
microscopic contact angle in the Young equation---requiring
$\inin{\rho}(\inin{r},\pi-\theta_S)=\ilfrac{1}{2}$, at least up to this
first-order correction, and leading to
\begin{equation*}
 \inin{\rho}_1 = -\frac{\sin\theta_S\cos\theta+\cos\theta_S\sin\theta}{4}, \
\lr{\begin{array}{c} \inin{u}_0 \\ \inin{v}_0 \end{array}} =
\lr{\begin{array}{c} -\cos\theta \\ \sin\theta \end{array}}.
\end{equation*}
We now consider the stresses and pressure, which in inner variables are
scaled with $\Cn^{-1}$ (readily seen from eqs.~\refe{eq:nd1}--\refe{eq:nd2}), as their singular behaviour in the classical
model of \cite{HuhScriv71} is the hallmark of the moving contact line problem. The total stress components in polar coordinates and in inner variables are
\begin{align*}
\inin{\tens M}_{{\inin{r}}{\inin{r}}}
&= {\Ca_k}^{-1}[
\{{\inin{\rho}^2(1-\inin{\rho})^2} +  \lr{\pilfrac{\inin{\rho}}{\theta}}^2/{\inin{r}^2} - \lr{\pilfrac{\inin{\rho}}{\inin{r}}}^2\} /2
\nonumber\\ &
+ \inin{\rho}\lrcur{\ppilfrac{\inin{\rho}}{\inin{r}} + \ilfrac{\pilfrac{\inin{\rho}}{\inin{r}}}{\inin{r}} + \ilfrac{\ppilfrac{\inin{\rho}}{\theta}}{\inin{r}^2}  - {\inin{\rho}(1-\inin{\rho})(1-2\inin{\rho})}}
]
\nonumber\\ &
+ \inin{\rho}\lrsq{{4}\pilfrac{\inin{u}}{\inin{r}}/{3}-{2}\lr{ \inin{u} + \pilfrac{\inin{v}}{\theta}}/{(3 \inin{r})}},
\\
\inin{\tens M}_{\inin{r}\theta}
&= -{\Ca_k}^{-1}\pilfrac{\inin{\rho}}{\inin{r}} \ \pilfrac{\inin{\rho}}{\theta}/\inin{r}
+ \inin{\rho}\lrsq{\lr{\pilfrac{\inin{u}}{\theta} -  \inin{v}}/\inin{r} + \pilfrac{\inin{v}}{\inin{r}}},
\\
\inin{\tens M}_{\theta\theta}
&=  {\Ca_k}^{-1}[
\{{\inin{\rho}^2(1-\inin{\rho})^2} -  \lr{\pilfrac{\inin{\rho}}{\theta}}^2/\inin{r}^2 + \lr{\pilfrac{\inin{\rho}}{\inin{r}}}^2\} /2
\nonumber\\ &
+ \inin{\rho}\lrcur{ \ppilfrac{\inin{\rho}}{\inin{r}} + \pilfrac{\inin{\rho}}{\inin{r}}/\inin{r} +  \ppilfrac{\inin{\rho}}{\theta}/{\inin{r}^2} - {\inin{\rho}(1-\inin{\rho})(1-2\inin{\rho})}}
]
\nonumber\\&
+ \inin{\rho}\lrsq{- {2}\pilfrac{ \inin{u}}{ \inin{r}}/3 + {4}\lr{ \inin{u} + \pilfrac{ \inin{v}}{\theta}}/{(3 \inin{r})}},
\end{align*}
so by substituting in our results, we find at leading order
$\inin{\tens{M}} = O(1)$, as all $O(1/\inin{r})$ terms cancel. To obtain the
precise form of the stresses, the second-order terms in the governing
equations are needed. The pressure in this inner region is given by $\inin p
= \inin\rho^2(1-3\inin\rho)(1-\inin\rho)/(2\Ca_k)$, so that as $\inin r \to 0$ it
satisfies $\inin p = -1/(32\Ca_k) + O(\inin r)$, being finite at the contact line.
Continuing with these second-order terms, we find the density and velocity
corrections
\begin{align*}
 \inin\rho_2 &= {C_{\inin \rho_1}+C_{\inin \rho_2}\cos(2\theta)}, \\
\lr{\begin{array}{c} \inin{u}_1 \\ \inin{v}_1 \end{array}} &=
\lr{\begin{array}{c}
{\sin\theta_S(\cos(2\theta)-1)/4 - C_{\inin v}\sin(2\theta)} \\
{-\sin\theta_S\sin(2\theta)/4 + C_{\inin v}(1-\cos(2\theta))}
\end{array}}
,
\end{align*}
arising through solving the governing equations
\begin{gather*}
 \inin{u}_1 = -\sin(\theta_S)/4 - \inin{v}_1'/2, \quad
 \inin{v}_1' = -\inin{v}_1'''/4, \\
 \inin{\rho}_2' = -\inin{\rho}_2'''/4 ,
\end{gather*}
and boundary conditions
\begin{align*}
 \inin{u}_1 = \inin{v}_1 = \inin{\rho}_2' = 0 \qquad \mbox{on }\theta=\{0,\pi\},
\end{align*}
at this order, and
where the arbitrary constants $C_{\inin \rho_1}$, $C_{\inin \rho_2}$ and
$C_{\inin v}$ would be set by the full solution of the inner problem. A
possible flow scenario where $C_{\inin \rho_1}=-0.1$, $C_{\inin \rho_2}=0.3$,
$C_{\inin v}=-1$, for $\theta_S=\pi/4$ is shown in Fig.~\ref{fig:fo_plots}(a). All of these results allow us to determine the
leading-order stress components as
\begin{align*}
 \inin{\tens M}_{\inin{r}\inin{r}} &=
 (M_1\cos(2\theta) -
 M_2\sin(2\theta) + M_3)/32 + O(\inin r),
 \\
 \inin{\tens M}_{\inin{r}\inin{\theta}} &=
 -(M_2\cos(2\theta) + M_1\sin(2\theta))/32 + O(\inin r),
 \\
 \inin{\tens M}_{\inin{\theta}\inin{\theta}} &= (M_2\sin(2\theta)
 -M_1\cos(2\theta) + M_3)/32 + O(\inin r),
\end{align*}
where 
\begin{align*}
 M_1 &= {\cos(2\theta_S)}/{\Ca_k}+8\sin\theta_S, \\
 M_2 &= 32C_{\inin v}+{\sin(2\theta_S)}/{\Ca_k}, \\
 M_3 &= {(1+64C_{\inin \rho_1})}/{\Ca_k}-\ilfrac{8\sin\theta_S}{3},
\end{align*}
showing that the stresses are
nonsingular as $\inin r \to 0$.

\section{Extensions} Having demonstrated the ability of our diffuse-interface model to alleviate the moving contact line
problem with no-slip applied, we now consider a number of other features of
contact line flow. These extensions are to demonstrate the range of boundary conditions derived and
applied in the literature, being relevant to various physical situations, and to draw comparisons between
diffuse-interface models and one of the more complex continuum theories proposed to deal with contact line flows, the interface formation model of Shikhmurzaev \cite{ShikhBook}.

A recent paper, \cite{My_Shikh}, critically examined this interface
formation model of Shikhmurzaev \cite{ShikhBook}.
There, it was shown that the model degenerates to the same macroscopic flow
as slip models but it was also seen to have features that most of the simpler
models such as Navier-slip do not.
In particular, the interface formation model captures finite pressure
behaviour, it generalises both the fluid-fluid and the fluid-solid interfaces
from their classical models of being sharp and with no-slip respectively
through the modelling of surface layers, and the microscopic contact angle is able to
vary dynamically from its static value (with its value determined as part of the solution
rather than prescribed empirically as it is sometimes for slip models, \eg~\cite{Hocking90}
). Finally, the fluid is able to `roll', as in a moving
frame of reference there is no stagnation point at the contact line, allowing particles
to reach and transfer through the contact line in finite time.
Whilst these features occur at lengthscales too small to probe with current
experimental ability (see discussions in \cite{My_Shikh}), it is of interest
that the diffuse-interface model is capable of similar predictions, with various features
added. The model studied thus far already predicts finite pressure at the contact line,
and alleviates the stagnation point predicted by slip models through mass
transfer. It relaxes the sharp fluid-fluid interface assumption, but has classical no-slip at
the wall in contrast to the effective slip of the interface formation model.
Although not necessary, this may also be relaxed through carefully
prescribing a generalised Navier boundary condition (GNBC), suggesting that
the slip velocity is proportional to the total tangential stress (the sum of
the viscous and uncompensated Young stress---arising from the deviation of
the fluid-fluid interface from its static shape), and derivable using
variational arguments from the principle of minimum energy dissipation
\cite{QianWangShengGNBCfirst,QianWangShengJFM}.
Our diffuse-interface model also prescribes the microscopic dynamic contact angle $\theta_d$ to be equal to the static value $\theta_S$ through the wetting boundary condition. An alternative is for this condition to hold at equilibrium, with the density relaxing to it in finite time when out of equilibrium, as initially discussed
(but not implemented) for binary fluids \cite{jacqmin}, and more recently used in numerical simulations \cite{QianWangShengGNBCfirst,QianWangShengJFM,YueFengEPJ,YueFengPoF}.

\begin{figure}[t]
\centering
	\includegraphics{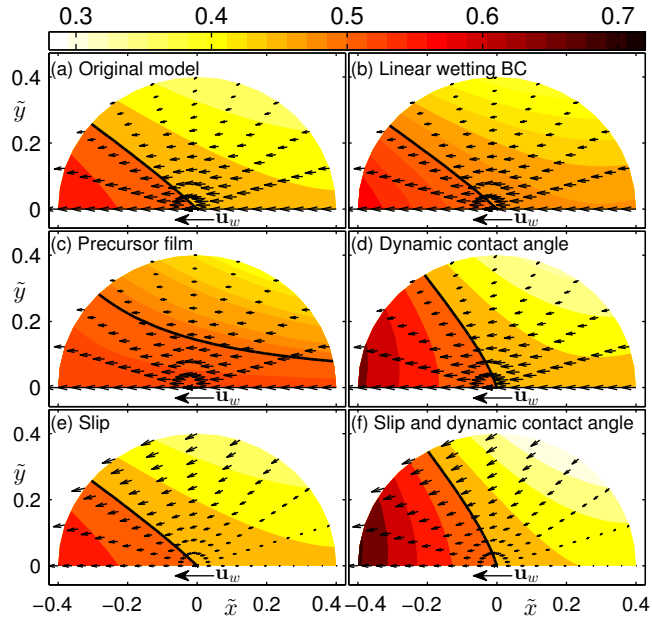}\	
	\caption{The asymptotic behaviour of density (contour plots) and velocity (arrows) as the contact line is approached. The driving force in the system is the moving wall.}
\label{fig:fo_plots}
\end{figure}

For our liquid-gas configuration, the GNBC (of Qian, Wang and Sheng \cite{QianWangShengGNBCfirst,QianWangShengJFM}) and generalised wetting boundary condition (of the variety of authors mentioned above \cite{jacqmin,QianWangShengGNBCfirst,QianWangShengJFM,YueFengEPJ,YueFengPoF}) may be considered analogously. In dimensional form the wetting boundary condition is generalised to
\begin{equation}
 \alpha\lr{\pilfrac{\bar\rho}{\bar t} + \bar\tbU\bmDot\bmN\bar\rho} = -\bar L(\bar\rho),\label{eq:wallrelaxwet}
\end{equation}
where $ \bar L(\bar\rho) = K \tbN_w \bmDot \bmN \bar\rho + \bar f'_w(\bar\rho)$, is the wall chemical potential, and
$\alpha=0$ representing instantaneous relaxation to equilibrium. A representative effect of \refe{eq:wallrelaxwet} is shown in Fig.~\ref{fig:fo_plots}(d), in comparison to \refe{eq:origbcs} in Fig.~\ref{fig:fo_plots}(a). The GNBC for this application with inverse slip length $\bar\beta$ is then
\begin{equation}
 \bar L(\bar\rho) (\vect{t}_w\bmDot\bmN\bar\rho) - \bar\bmtau_{nt} = \bar\beta(\bar\tbU-\bar\tbU_w)\bmDot\vect{t}_w,
\label{eq:dimGNBC}
\end{equation}
with $\vect{t}_w$ the tangent to the wall, and $\bar\bmtau_{nt}$ the viscous shear stress. Note $\alpha=0$ reduces to the popular Navier-slip condition. A representative effect of the GNBC from \refe{eq:dimGNBC} is shown in Fig.~\ref{fig:fo_plots}(e), and in combination with \refe{eq:wallrelaxwet} in Fig.~\ref{fig:fo_plots}(f).
In nondimensional form in polar coordinates for steady flow in the inner region, \refe{eq:wallrelaxwet} reduces to
\begin{equation}
 \inin L(\inin\rho) = \Ca_k \Pi \lr{\inin u \, \pilfrac{\inin\rho}{\inin r} +
 {\inin v} \, \pilfrac{\inin\rho}{\theta}/{\inin r}},\label{eq:wallrelaxwetin}
\end{equation}
where $\inin L(\inin\rho) = \pm\pilfrac{\inin\rho}{\theta} /\inin r + \inin\rho(1-\inin\rho)\cos\theta_S$ on $\theta=\{0,\pi\}$, and with another nondimensional parameter arising, $\Pi={\alpha\rho_L^2}/({\mu_L\epsilon})$, describing the extent of the wall relaxation ($\Pi=0$ being instantaneous). Similarly \refe{eq:dimGNBC} reduces to
\begin{equation}
 \mp \inin L(\inin\rho)\pilfrac{\inin\rho}{\inin r}/\Ca_k
 - \inin\rho\lrsq{(\pilfrac{\inin u}{\theta} - {\inin v})/{\inin r} + \pilfrac{\inin v}{\inin r}} = \beta (1\pm \inin u),\label{eq:dimGNBCin}
\end{equation}
where $\beta = {\bar\beta \epsilon}/{\mu_L}$ is the nondimensional slip parameter, and along with $\Pi$ are both chosen to be formed with the interface thickness $\epsilon$ such that they are considered as $O(1)$ in the limit $\Cn\to0$.

To consider how \refe{eq:wallrelaxwet} allows for microscopic contact angle
variation dependent on flow conditions, we note that \refe{eq:wallrelaxwet} at equilibrium (denoted
with subscript $e$) gives 
\begin{equation*}
 \bar L(\bar\rho) =
\lreval{K \tbN_w \bmDot \bmN \bar\rho}_e + \bar f'_w(\rho_L/2) = 0.
\end{equation*}
Based on calculations
in \cite{QianWangShengJFM,YueFengPoF} for the binary fluid case,
we consider a
steady, dynamic situation (denoted with subscript $d$), and see that
\refe{eq:wallrelaxwet} implies
\begin{equation*}
 \alpha
\lreval{\bar\tbU\bmDot\bmN\bar\rho}_{d} = - \lreval{K \tbN_w \bmDot \bmN
\bar\rho}_{d} - \bar f'_w(\rho_L/2),
\end{equation*}
thus using the equilibrium result and
considering this at the contact line with wall velocity $\bar\tbU_w=-V\vect{t}_w$, we
find $ V\alpha\sin\theta_d = - K (\cos\theta_d - \cos\theta_S)$, or in
nondimensional form
\begin{equation*}
 \Ca_k\Pi = \ilfrac{(\cos\theta_S- \cos\theta_d)}{\sin\theta_d} \approx \theta_d-\theta_S,
\end{equation*}
where the final approximation holds for $\theta_d -\theta_S \ll 1$, in agreement with the binary fluid case \cite{YueFengPoF}.

Another consideration is the behaviour near the contact line if density gradients near the wall are permitted far from the contact line. Our wall free energy was specifically chosen to prevent this, but we may also consider the two other situations used previously in the literature, namely (i) specifying a density at the wall $\bar\rho=\rho_a$ on $\partial\Omega$, as in \cite{PismenPomeau}, and a representative effect of this shown in Fig.~\ref{fig:fo_plots}(c) and (ii) choosing a linear form in the density for the wall free energy $\bar f_w(\bar\rho) = a\bar\rho$, as in \cite{seppecher,BriantYeomansEarly,BriantYeomans1,qianliqgas}, and shown in Fig.~\ref{fig:fo_plots}(b). To consider the contact line behaviour when (i) replaces the wetting boundary condition is straightforward, but for (ii), we must understand how to impose the microscopic contact angle to compare to our previous condition. Following \cite{BriantYeomansEarly}, we use Young's law $\cos\theta_S = (\sigma_{V}-\sigma_{L})/\sigma$ and compute $\sigma_{V}$ and $\sigma_{L}$ by integrating the free energy per unit area along the corresponding interface. This gives $\cos\theta_S = [(1-A)^{3/2}-(1+A)^{3/2}]/2$, where $A = 4a\epsilon/(K\rho_L)$ is nondimensional. This may then be inverted to give the appropriate value of $A$ for a given contact angle $\theta_S$, and corresponds to the nondimensional boundary condition
$ \Cn \tbN_w \bmDot \bmN \rho = -A/4 $.

Adding these features into the diffuse-interface model do not dramatically alter the contact line behaviour, but subtle differences in the asymptotic results are demonstrated, as mentioned, in Fig.~\ref{fig:fo_plots} for selected arbitrary constants, where $\Ca_k=0.1$ and $\theta_S=\pi/4$, and may be compared to the equilibrium situation in Fig.~\ref{fig:eqd}. The cases considered are (a) the original model \refe{eq:origbcs} without slip or wall relaxation, (b) using the linear form for
$\bar{f}_w(\bar\rho)$
in the wetting boundary condition, (c) adding a precursor film at the wall (where the wetting boundary condition is replaced by $\rho=\rho_a/\rho_L=0.53$), (d) allowing finite wall relaxation, \refe{eq:wallrelaxwetin} with $\Pi=5$, (e) including the GNBC (\refe{eq:dimGNBCin} with $\beta=2$) but with $\Pi=0$, and (f) using \refe{eq:wallrelaxwetin} with $\Pi=5$ and \refe{eq:dimGNBCin} with $\beta=2$. All models behave as expected, resolving the stress and pressure singularities, and including the effects they intend, \eg~capturing the film in (c), increased microscopic contact angle in (d) and (f), and reduced wall velocity in (e) and (f). There are only small differences between cubic and linear wall energy forms near the contact line, mainly that the linear form shows a broader band of density variation. This hints at the important difference that will occur near the wall but far away from the interface, where density gradients will remain present for this linear form.

\section{Conclusions} We have shown analytically that a diffuse-interface model is able to resolve the moving contact line
problem through relaxing the interface from being sharp to thin, without need
to model any further physical effects from the microscale at the contact line. Whilst slip, precursor films
and finite-time wall relaxation have been considered, they are \emph{not}
necessary to resolve the moving contact line problem. We believe that the present study will
motivate further analytical and numerical work with diffuse-interface models, such as to
consider heterogeneous walls
\cite{Savva09,Savva10,RajChemHet,christophe_marc}. Of particular interest
would also be the inclusion of non-local terms into the governing equations;
this was considered in \cite{antoniojfm} for equilibrium wetting using a
density-functional theory.

\section*{Acknowledgements}
We acknowledge financial support from ERC Advanced Grant No. 247031, and Imperial College through a DTG International Studentship.

\bibliographystyle{abbrv}
\bibliography{arxiv_epje_CHbib}

\end{document}